\documentclass[prd,nofootinbib,preprintnumbers,twocolumn
]{revtex4}
\usepackage{graphicx,epsf,epsfig}
\usepackage{amsmath}
\usepackage{amssymb}
\usepackage{bbm}
\newcommand{\cm}{{\rm c}}
\newcommand{\sm}{{\rm s}}

\usepackage{tabls,multirow}

\def\eqs#1#2{{eqs.~(\ref{#1})--(\ref{#2})}}

\def\hbar{\hspace{0pt}\raisebox{1pt}{$-$} \hspace{-7pt} h}

\newcommand{\be}{\begin{equation}}
\newcommand{\ee}{\end{equation}}
\newcommand{\bd}{\begin{displaymath}}
\newcommand{\ed}{\end{displaymath}}
\newcommand{\bea}{\begin{eqnarray}}
\newcommand{\eea}{\end{eqnarray}}
\newcommand{\nn}{\nonumber}

\def\so10{$SO(10)$}

\begin{document}
\title{Quark and lepton masses and mixing in \so10 with a GUT-scale vector matter}
\date{December 9, 2007}
\author{Michal Malinsk\'{y}}\email{malinsky@phys.soton.ac.uk}
\affiliation{School of Physics and Astronomy, University of Southampton,
SO16 1BJ Southampton, United Kingdom}
\begin{abstract}
We explore in detail the effective matter fermion mass sum-rules in a class of renormalizable SUSY $SO(10)$ grand unified models where the quark and lepton mass and mixing patterns originate from non-decoupling effects of an extra vector matter multiplet living around the unification scale.
If the renormalizable type-II contribution governed by the $SU(2)_{L}$-triplet in $54_{H}$ dominates the seesaw formula, we obtain an interesting correlation between the maximality of the atmospheric neutrino mixing and the proximity of $y_{s}/y_{b}$ to $V_{cb}$ in the quark sector.
\end{abstract}
\maketitle
\section{Introduction}
Though being on the market for more than thirty years, the idea of grand unification \cite{Georgi:1974sy} still receives a lot of attention in various contexts.  Not only it is physical~-  the proton decay, one of its inherent consequences, is experimentally testable - but it has also been very successful in shedding light on some of the deepest mysteries of the Standard Model (SM), be it the (hyper)charge quantization, the electroweak scale gauge coupling hierarchy, or the high scale $b-\tau$ Yukawa coupling convergence.

The advent of precision neutrino physics in the last decade triggered an enormous boost 
to the field. The unprecedented smallness of the neutrino masses \cite{Strumia:2006db}, finding a natural explanation in the class of seesaw models \cite{seesaw1}, indicates a new fundamental scale at around $10^{14}$~GeV, remarkably close to the gauge coupling unification scale of simplest grand unifications (GUTs) $M_{\mathrm{GUT}}\approx 2\times 10^{16}$ GeV.    
Moreover, the high degree of complementarity among the forthcoming large volume neutrino experiments and proton decay searches \cite{protondecaysearches1}  
is promising a further GUT renaissance in the near future.

An important new aspect of the recent developments is the proliferation of detailed studies of the Yukawa sector \cite{Yukawastudies1} of various GUTs including reliable data on lepton mixing, with a potential to further constrain the simplest predictive models like e.g. the minimal supersymmetric $SU(5)$ \cite{Dimopoulos:1981zb} or $SO(10)$ \cite{MinimalSO10}. This, in turn, has nontrivial implications for proton decay \cite{protondecay3}, absolute neutrino mass scale \cite{absoluteneutrinomassscale}, leptogenesis \cite{SO10leptogenesis} etc. 

In what follows, we shall stick to the class of $SO(10)$-based supersymmetric (SUSY) GUTs. 
Perhaps the most attractive feature of the $SO(10)$ schemes is their capability of accommodating all the SM matter multiplets within just three $16$-dimensional spinor representations, thus providing a simple understanding of the peculiar SM hypercharge pattern. 
Moreover, the proton decay issue of the minimal SUSY $SU(5)$ can be considerably alleviated in SUSY SO(10), see e.g. \cite{Babu:1993we,SO10protondecay}, because $SO(10)$, as a rank 5 group, features higher flexibility in the symmetry breaking pattern \cite{SO10breakingpattern, SO10breakingpattern2,Aulakh:2000sn} than $SU(5)$.  

Concerning the basic symmetry breaking scenarios, the two most popular variants are distinguished by either making use of 16-dimensional spinors, or the 252-dimensional 5-index antisymmetric tensor (decomposing under parity into $126\oplus \overline{126}$ self- and anti-selfdual components) in the Higgs sector in order to break the intermediate $SU(2)_{R}\otimes U(1)_{B-L}$ symmetry downto $U(1)_{Y}$ of the SM hypercharge. However, as the SM singlets in both $16_{H}\oplus \overline{16}_{H}$ and $126_{H}\oplus \overline{126}_{H}$ preserve $SU(5)$, extra Higgs multiplets like $45_{H}\oplus 54_{H}$ or $210_{H}$ are needed (c.f. \cite{MinimalSO10, SO10breakingpattern2,Aulakh:2000sn} and references therein) to achieve the full $SO(10)\to$ SM gauge symmetry breakdown. On top of that, the correlations between the effective Yukawa couplings strongly suggest an extra $10_{H}$ in the Higgs sector taking part at the final SM\nolinebreak $\to SU(3)_{c}\otimes U(1)_{Q}$ step.

Though similar in the symmetry breaking strategy, these options differ dramatically at the effective Yukawa sector level. Concerning the $\overline{126}_{H}\oplus 126_{H}$ case, the anti-selfdual part $\overline{126}_{H}$ couples to the spinorial matter bilinear $16_{F} 16_{F}$ via renormalizable coupling $16_{F}16_{F}\overline{126}_{H}$, which (together with the  $16_{F}16_{F}10_{H}$ vertex) gives rise to simple effective Yukawa and Majorana sector sum-rules featuring a high degree of predictivity in the matter sector \cite{MinimalSO10,Yukawastudies4,SO10protondecay, Yukawastudies2}. 

The minimal potentially realistic renormalizable scenario of this kind (MSGUT) \cite{MinimalSO10} (with $10_{H}$,  $126_{H}\oplus \overline{126}_{H}$ and $210_{H}$ in the Higgs sector
) became a subject of thorough examination in the past few years \cite{b-taularge23, MinimalSO10examination,Yukawastudies2,Yukawastudies3,absoluteneutrinomassscale}. This was triggered namely by the observation \cite{b-taularge23} of a profound link between the large 23-mixing in the lepton sector and the (GUT-scale) $b-\tau$ Yukawa convergence, if the type-II contribution (coming from a renormalizable coupling $16_{F}16_{F}\overline{126}_{H}$) governs the seesaw formula. It has also been shown \cite{b-taularge23,Yukawastudies3} that this scenario predicts a  relatively large reactor mixing angle (typically $\sin \theta_{13}\approx 0.1$ ), well within the reach of the future neutrino experiments \cite{Ue3measurements}.  However, the recent studies \cite{absoluteneutrinomassscale} revealed a tension between the lower bounds on the absolute neutrino mass scale and the GUT-scale gauge coupling unification thresholds.  

Though these issues can be to some extent relaxed by adding an extra $120$-dimensional Higgs multiplet,
either as a subleading correction to the minimal model setting \cite{adding120} or as a full-featured contribution to the relevant Yukawa sum-rules (deferring $\overline{126}_{H}$ for the neutrino sector purposes \cite{NMSGUT}), the large Higgs sector generically pulls the Landau pole to the GUT scale proximity \cite{Aulakh:2000sn}, thus questioning the viability of the perturbative approach.  

If, on the other hand, $16_{H}\oplus \overline{16}_{H}$ is employed\cite{Babu:1993we,simpleHiggssector1}, the complexity of the Higgs sector is reduced and the Landau pole issue can be partially relaxed.
With just three matter spinors, the renormalizable operators are incapable of transferring the effects of $SU(2)_{R}\otimes U(1)_{B-L}$ breaking into the matter sector and thus non-renormalizable couplings must be invoked. This, however, ruins the Yukawa sector predictivity of such theories, unless extra assumptions (like e.g. family symmetries) reduce the number of free parameters entering the effective mass matrices, see e.g. \cite{Yukawastudies1}. 

A simple way out \cite{vectormatter} consists in adding extra matter multiplets, in particular the 10-dimensional $SO(10)$ vector(s) transmitting the $SU(2)_{R}\otimes U(1)_{B-L}$ breakdown (driven by $16_{H}\oplus\overline{16}_{H}$) to the effective matter sector (spanning over $16_{F}^{i}\oplus 10_{F}$) via renormalizable mixing terms $16_{F}^{i}10_{F}16_{H}$. Though $10_{F}$ tends to decouple from the GUT-scale physics upon pushing its $SO(10)$-singlet mass $M_{10}$ far above the GUT scale, the generic proximity of the above-GUT-scale thresholds (being it $M_{\mathrm{Planck}}$ or just a higher unification scale) admits for speculation about the lightest such multiplet next to the GUT-scale. 

With $10_{F}$ at hand, the $SU(5)$ breaking (triggered typically by $45_{H}\oplus 54_{H}$) can also be transmitted to the matter sector via loops (in non-SUSY context), higher order operators \cite{Barr:2007ma}, or at renormalizable level via $10_{F}10_{F}54_{H}$ or\footnote{Recall that due to antisymmetry of $45_{H}$ the latter option is viable only with more than one $10_{F}$.} $10_{F}10_{F}45_{H}$  couplings. 
Although there is a number of studies exploiting this mechanism in the literature \cite{vectormatter,vectormatterused}, a generic viability analysis of this strategy is still missing. 

In this paper, we shall attempt to fill this gap by focusing on the simplest such scenario, a renormalizable 
model with three matter spinors $16_{F}^{i}$ (for $i=1,2,3$) and one extra vector multiplet $10_{F}$.  
We will not resort to any extra symmetries or effective operators or make other assumptions to reduce the complexity of the effective Yukawa sum-rules; rather than that we shall scrutinize the {\it generic} setting analytically (focusing in particular on the quark and charged lepton sector) 
in order  to get as much understanding of the numerical results as possible.
The neutrino sector does not admit for such a detailed analysis unless the type-II contribution happens to govern the seesaw formula. In such a case, one of the lepton sector mixing angles is under control and one can derive a new (GUT-scale) relation between the deviation of the lepton sector 23 mixing from maximality and the proximity of $y_{s}/y_{b}$ and $V_{cb}$ in the quark sector.

The paper is organized as follows: in section \ref{sectmodel}, the relevant SUSY $SO(10)$ framework is defined, with particular attention paid to the generic features of the Yukawa sector. Next, we derive a detailed form of the relevant GUT-scale Yukawa matrices and (after integrating out the heavy degrees of freedom) give the effective 3$\times$3 mass matrices for the SM matter fermions. Section IV is devoted to a thorough analysis of these structures, the relevant parameter counting and development of tools necessary for analytic understanding of the given numerical results (deferring some of the technicalities into an Appendix). Focusing on the heavy sector, we shall examine the correlation between the proximity of $y_{s}/y_{b}$ to $V_{cb}$ and the large atmospheric lepton mixing.   
\section{The model\label{sectmodel}}
Following the strategy sketched above, the matter sector of the scenario of  our interest consists of three 'standard' copies of the $SO(10)$ spinorial matter
residing in $16_F^i$ ($i=1,2,3$) plus one extra
$SO(10)$ vector representation
$10_F$. 
The $SU(3)_{c}\otimes SU(2)_{L}\otimes U(1)_{Y}$ decomposition of these
multiplets (up to the generation indices) reads:
\bea\label{decs} 
16_F & =& (3,2,+\tfrac{1}{3})\oplus(1,2,-1)\oplus(\overline{3},1,-\tfrac{4}{3}) \\
& \oplus & (\overline{3},1,+\tfrac{2}{3})\oplus(1,1,0)\oplus(1,1,+2)\nn\\
10_F & = & (3,1,-\tfrac{2}{3})\oplus(1,2,+1)\oplus (\overline{3},1,+\tfrac{2}{3})\oplus(1,2,-1)\nn
\eea
As usual, the sub-multiplets of $16_{F}$ will be consecutively referred to as  $Q_L$, $L_L$,
$U^c_L$, $D^c_L$, $N^c_L$  and  $E^c_L$, while those in the decomposition of $10_{F}$ as
 $\Delta_L$, $\Lambda^c_L$, $\Delta^c_L$ and  $\Lambda_L$.
Therefore, at the SM level, $D^{c}_{L}$ can mix with $\Delta^{c}_{L}$ and $L_{L}$ with $\Lambda_{L}$ giving rise to the physical light down-type quark and charged-lepton mass eigenstates  $d_{L}^{c}$ and $l_{L}$ (which then share the features of both $16_{F}^{i}$ and $10_{F}$, in particular the sensitivity of $10_{F}$ to the $SU(2)_{R}\otimes U(1)_{B-L}$ and $SU(5)$ breakdown).

The Higgs sector is taken to be the minimal (concerning dimensionality) leading to a viable symmetry breaking chain (while preserving SUSY down to the electroweak scale), i.e. $10_{H}\oplus 16_{H}\oplus\overline{16}_{H}\oplus 45_{H}\oplus 54_{H}$ \cite{Aulakh:2000sn}. We shall further assume an unbroken $Z_{2}$ parity distinguishing among the ($Z_{2}$-odd) matter multiplets $16_{F}$ and $10_{F}$  and the Higgs sector fields (that are $Z_{2}$-even). 
The relevant SM decompositions read:
\bea
10_H & = & (3,1,-\tfrac{2}{3})  \oplus  \underline{(1,2,+1)}\oplus (\overline{3},1,+\tfrac{2}{3})\oplus\underline{(1,2,-1)}\nn
\\
16_H & =& (3,2,+\tfrac{1}{3})\oplus\underline{(1,2,-1)}\oplus(\overline{3},1,-\tfrac{4}{3})\nn \\
& \oplus & (\overline{3},1,+\tfrac{2}{3})\oplus(1,1,+2)\oplus\underline{(1,1,0)}\nn
\\
\overline{16}_H & =& (\overline{3},2,-\tfrac{1}{3})\oplus\underline{(1,2,+1)}\oplus(3,1,+\tfrac{4}{3})\nn \\
& \oplus & (3,1,-\tfrac{2}{3})\oplus(1,1,-2)\oplus\underline{(1,1,0)}
\\
54_H & = & \underline{(1,1,0)}\oplus(1,3,0)\oplus(1,3,\pm 2)\nn \\
&\oplus& (\overline{6},1,+\tfrac{4}{3})\oplus(6,1,-\tfrac{4}{3})\oplus(8,1,0)\oplus (3,2,+\tfrac{1}{3})\nn \\
&\oplus& (3,2,-\tfrac{5}{3}) \oplus (\overline{3},2,-\tfrac{1}{3})\oplus(\overline{3},2,+\tfrac{5}{3})\nn
\eea
The underlined components of the SM singlet type in $16_{H}\oplus\overline{16}_{H}$ and $54_{H}$ receive GUT-scale
VEVs while the doublets $(1,2,\pm 1)$ enter the light $SU(2)_{L}$-doublets responsible for the
electroweak symmetry breakdown. We shall use:
$H_u^{10}\equiv (1,2,+1)_{10}$, $H_d^{10}\equiv (1,2,-1)_{10}$,
$H_u^{\overline{16}}\equiv (1,2,+1)_{\overline{16}}$, $H_d^{16}\equiv (1,2,-1)_{16}$, 
$S^{16}\equiv (1,1,0)_{16}$, $S^{\overline{16}}\equiv
(1,1,0)_{\overline{16}}$ and $S^{54}\equiv (1,1,0)_{54}$ and 
$\langle S^{16} \rangle\equiv V^{16}$, $\langle S^{\overline{16}} \rangle\equiv V^{\overline{16}}=(V^{16})^{*}$ (from $D$-flattness), $\langle S^{54}\rangle\equiv V^{54}$, $\langle H_{u,d}^{10}\rangle \equiv v_{u,d}^{10}$, $\langle H_{d}^{16}\rangle \equiv v_{d}^{16}$ for the corresponding VEVs.
 
Though $10_{F}$ (unlike $16_{F}^{i}$) admits an $SO(10)$ singlet mass term $M_{10}10_{F}10_{F}$ in the superpotential  and thus should decouple in the $M_{10}\to \infty$ limit \cite{decoupling}, we shall assume the opposite, i.e. that $M_{10}$ happens 
 to live close to the $SO(10)$ breaking scales $V^{54}$ and $V^{16}$. In such a case, the effective light matter becomes sensitive to the GUT-symmetry breakdown due to the interactions of its non-vanishing components in the $10_{F}$ direction with the $SU(2)_{R}\otimes U(1)_{B-L}$ breaking VEVs in $16_{H}$ (via $10_{F}16_{F}16_{H}$) and the $SU(5)$ breaking VEVs in $45_{H}\oplus 54_{H}$ (through $10_{F}10_{F}54_{H}$). 

For this to be the case, one must assume that the mixing terms $16_{F}^{i}10_{F}16_{H}$ not to be very suppressed with respect to $M_{10}$ and $V^{54}$ (driving the mass of the heavy part of the matter sector), which, however, is exactly the situation suggested by the gauge-coupling renormalization group running in the SUSY 'desert' picture.

\subsection{The superpotential}
The most general renormalizable (and $Z_{2}$-even) Yukawa superpotential at the $SO(10)$ level reads
\bea
W_Y & = & Y^{ij}16^i_F 16^j_F 10_H+F^i 16^i_F 10_F 16_H \nn\\
\label{WY}
&+&\lambda 10_F 10_F 54_H + M_{10}10_F 10_F.
\eea
Here $Y$ denotes a $3\times 3$ symmetric complex Yukawa matrix, while $\vec{F}$ and $\lambda$ are the relevant 3-component complex vector and scalar Yukawa couplings respectively.
In components, this gives rise to the following $SU(3)_c\otimes SU(2)_L \otimes U(1)_Y$ structure (in the LR chirality basis):
\bea
W_Y&\ni& Y^{ij}\left[Q_L^i U^{cj}_L H_u^{10}+Q_L^i D^{cj}_L H_d^{10}+L_L^i N^{cj}_L H_u^{10}\right.\nn \\
&+&\left.L_L^i E^{cj}_L H_d^{10}\right]+F^i\left[L_L^i \Lambda_L^c S^{16}-D^{ci}_L\Delta_L S^{16}\right.\nn\\
&+&Q_L^i \Delta_L^c H_d^{16}-\left.E^{ci}_L\Lambda_L H_d^{16}+N^{ci}_L\Lambda^c_L H_d^{16}\right]\nn\\
&+&\lambda\left[\Delta_L\Delta_L^c\left(M_{10}-\tfrac{1}{\sqrt{15}}S^{54}\right)\right.\label{WY2}\\
&+&\left.\Lambda^c_L\Lambda_L\left(M_{10}+\tfrac{1}{2}\sqrt{\tfrac{3}{5}}S^{54}\right)\right]+{\rm transposed}\nn\\
&+&\lambda\, c_{T}\left[\Lambda_L^c\Lambda_L^c T_-^{54}
+\Lambda_L\Lambda_L T_+^{54}\right]+\ldots \nn
\eea
where ``+transposed'' stands for the hermitean-conjugated terms (in the SUSY notation) while
$T_\pm^{54}\equiv (1,3,\pm 2)_{54}$ are $SU(2)_L$ triplets that can receive tiny VEVs relevant for the type-II neutrino mass matrix entry and $c_{T}$ is the associated Clebsch-Gordon (CG) coefficient.
\subsection{The GUT-scale mass matrices\label{GUTscalematrices}}
Once the relevant Higgs fields develop the VEVs specified above, 
the Yukawa couplings in $W_{Y}$ give rise to the quark and lepton mass matrices.
\paragraph*{Up-type quarks:}
The up-type quark mass matrix receives a simple $3\times 3$ form as there is no $(3,2,+\tfrac{1}{3})$-type multiplet in $10_{F}$ that could mix with the three up-quark-states in $16_{F}^{i}$:
\be\label{Muproto}
M_u = Y v_u^{10}.
\ee
\paragraph*{Down-type quarks:}
Since the right-handed down-quark-type states  $D^{c}_{L}$ in $16_F^{i}$ mix with $\Delta^{c}_{L}$ in $10_F$, the relevant (GUT-scale) mass-matrix is four-dimensional and reads (in the $\{D_{L}^{i},\Delta_{L}\}$ basis):
\be\label{Mdproto}
M_{d}=\left(
\begin{array}{cc}
Y v_d^{10} &- \vec{F} v_d^{16}\\
\vec{F}^T V^{16} & M_{10}-\lambda \tilde{V}^{54}
\end{array}\right).
\ee
The minus signs in the 4th column come from the relevant CG coefficients in (\ref{WY2}) with redefinition of $\tilde{V}^{54}\equiv {\tfrac{1}{\sqrt{15}}}V^{54}$.

\paragraph*{Charged leptons:}
The situation in the charged lepton sector is similar to the down-quark case up to the point that it is now the {\it left-handed} chiral components ($L_{L}$ in $16_{F}^{i}$ and  $\Lambda_{L}$ of $10_F$) that can mix. The net effect of this difference boils down to the charged lepton mass matrix structure very close to the transpose of $M_{d}$ (in the $\{E_{L}^{i},\Lambda_{L}^{-}\}$ basis, where $\Lambda_L^-$ denotes the charged component of the $\Lambda_L$ $SU(2)_{L}$-doublet):
\be\label{Mlproto}
M_{l}=\left(
\begin{array}{cc}
Y v_d^{10} & \vec{F} V^{16}\\
 -\vec{F}^T v_d^{16} & M_{10}+\frac{3}{2}\lambda \tilde{V}^{54}
\end{array}
\right).
\ee
Thus, it is namely the difference in the 44-entry CG coefficient that actually makes $M_{l}$ and $M_{d}$ feel the $SU(5)$ breakdown.
Moreover, this is also the {\it only} distinction between $M_{d}$ and $M_{l}^{T}$ in the current model and one of our goals will be to see whether such a detail could account for all the difference amongst the charged lepton and down quark spectra.

\paragraph*{Neutrinos:}
Since there are in total 8 neutral components in $16_F^i$ and $10_F$ ($N_L^i$, $N_L^{ci}$, $\Lambda_L^{c0}$ and $\Lambda_L^{0}$), the (symmetric) renormalizable neutrino mass matrix (in the $\{N_L^{i},N^{ci}_L,\Lambda_L^{0},\Lambda_L^{c0}\}$ basis) is more complicated:
\be
M_{\nu}=\left(
\begin{array}{cccc}
0 &  Y v_u^{10} & 0 & \vec{F} V^{16} \\
. & 0 & 0 & \vec{F} v_d^{16} \\
 . &  . & \lambda\, c_{T}\, w_+ & M_{10}+ \frac{3}{2}\lambda \tilde{V}^{54}\\
 . & . & . & \lambda\, c_{T}\, w_-
\end{array}\right).
\ee
Here we use $w_{\pm}$ for the VEVs of the electroweak triplets  $T_\pm^{54}=(1,3,\pm2)_{54}$, which provide the only source of  diagonal Majorana masses at the renormalizable level.

It is clear that this basic texture can not accommodate the standard seesaw mechanism: the 1-3 rotation, which cancels the large 14 entry (so that all the GUT-scale masses occupy the 3-4 sector), affects only the 11 entry of the 1-2 block (due to the zeros at the 13, 23 and 31, 32 positions). This gives rise to a type-II contribution proportional to $\lambda\, c_{T} w_{+}$ at 11-position, while keeps the other 1-2 block entries intact. We are then left with pseudo-Dirac neutrinos around the electroweak scale, at odds with experiment. 

However, the picture changes dramatically beyond the renormalizable level. The SM-singlet zero at the 22 position is not protected by the $SU(2)_{L}\otimes U(1)_{Y}$ symmetry and thus receives   contributions from the effective operators of the form $\kappa^{ij}16^{i}_{F}16^{j}_{F}\overline{16}_{H}\overline{16}_{H}/\Lambda_{\kappa}$ and $\tilde{\kappa}^{ij}16^{i}_{F}16^{j}_{F}16_{H}16_{H}/\Lambda_{\kappa}$ (with $\Lambda_{\kappa}$ denoting the relevant physics scale above $M_{\mathrm{GUT}}$) leading to a naturally suppressed\footnote{with respect to the typical GUT-scale $M_{M}$ generated at renormalizable level in models with $\overline{126}_{H}$ in the Higgs sector} Majorana mass term $M_{M}$. Next, the zero at the 23 position can be lifted upon the SM symmetry breakdown (requiring only one VEV insertion for the hypercharge deficit $+1$) by means of an effective operator  with structure $\zeta^{i}16^{i}_{F}10_{F}\overline{16}_{H}10_{H}/\Lambda_{\zeta}$. Last, a pair of electroweak VEV insertions (coming from the two types of the  effective operators above, plus $\sigma 10_{F}10_{F}{10}_{H}10_{H}/\Lambda_{\sigma}$), can give rise to the  $Y_{W}=+2$ entries at the doublet-doublet positions 11,13 and 33. 

With all this at hand, one finds that the only terms that do affect the structure of the seesaw formula are indeed the large Majorana mass $M_{M}\equiv \kappa (V^{\overline{16}})^{2}/\Lambda_{\kappa}$ at the 22 position, the electroweak VEVs at the 23/32 positions and the tiny diagonal Majorana VEVs $\mu_{3}\propto \kappa (v_{u}^{\overline{16}})^{2}/\Lambda_{\kappa}$ and $\mu_{\pm}\equiv \lambda c_{T}w_\pm+{\cal O}[(v_{u}^{10})^{2}/{\Lambda_{\sigma}}]$.  The neutrino mass matrix beyond the renormalizable level then reads (in the $\{N_L^{i},N^{ci}_L,\Lambda_L^{0},\Lambda_L^{c0}\}$ basis) at the leading order :
\be
M_{\nu}=
\left(
\begin{array}{cccc}
\mu_{3} &  Y v_u^{10} &. & F V^{16}  \\
 . & M_{M} & \vec{\zeta}\, V^{\overline{16}}v_{u}^{10}/\Lambda_{\zeta}  & F v_d^{16}\\
 . &  . & \mu_+ & M_{10}+ \frac{3}{2}\lambda \tilde{V}^{54}\\
 . & . & . & \mu_{-}
\end{array}
\right)
\ee
It is obvious that, without extra assumptions, $M_{\nu}$ is not constrained enough to admit for predictions in the neutrino sector (due to the ambiguity in the $M_{M}$ and $\mu_{3}$ matrices generated at the effective operator level only).
However, as we shall see in the next section, the renormalizable part of the effective type-II contribution to the effective light neutrino mass matrix (coming from the $SU(2)_{L}$-triplet of $54_{H}$) is calculable
up to an overall scale, and if it happens to dominate the seesaw formula, one can obtain an interesting link between the 23-quark sector observables and a large value of the associated lepton sector mixing!

\section{Effective mass sum-rules}
Since the up-quark mass matrix (\ref{Muproto}) is simple, let us focus on the down-quark and charged lepton mass matrices given by (\ref{Mdproto}) and (\ref{Mlproto}). 
First, in order to employ the CG coefficients at the 44 positions of $M_{d}$ and $M_{l}$ (to disentangle their effective light spectra), $10_{F}$ should not decouple. Therefore, we need $M_{10}$ comparable to  $V^{54}$  and $V^{16}$ not far below $V^{54}$; a detailed discussion can be found in section \ref{physics}. Thus, the physically viable situation corresponds to $V^{16}\lesssim V^{54}\approx M_{10}$ and from now on we shall always assume this to be the case.    
\subsection{Integrating out the heavy degrees of freedom\label{intout}}
With one dominant column in $M_{d}$ and $M_{l}^{T}$, there will always be three light matter states and one superheavy living around the GUT scale. If $V^{16} < V^{54}\approx M_{10}$, the light states are predominantly spanned over the $16_{F}^{i}$ components with a subleading contribution in the $10_{F}$ direction, while there is no such a clear mapping if $V^{16}\approx V^{54}\approx M_{10}$. Prior getting to the quantitative analysis of the effective SM spectra and mixings, the heavy degrees of freedom must be integrated out.
\paragraph*{Down-type quarks:} 
In the down-quark sector, 
the right-handed (RH) part of the physical heavy state ($\tilde{\Delta}$) can be readily identified:
\bea\label{hDelta}
\tilde\Delta^{c}_{L}&=&{\tfrac{1}{M_{\Delta}}}[F^{i} V^{16} D^{ci}_{L}+(M_{10}-\lambda\tilde{V}^{54}) \Delta_{L}^{c}]\nn\\
&\equiv& C_{d}^{i} D^{ci}_{L} + D_{d} \Delta_{L}^{c}
\eea
where $M_{\Delta}=\sqrt{F^{\dagger}F (V^{16})^{2}+(M_{10}-\lambda\tilde{V}^{54})^{2}}$ is a real normalization factor and 
\be\label{downparams}
\vec{C}_{d}=\vec{F}V^{16}/M_{\Delta},\quad 
{D}_{d}=(M_{10}-\lambda\tilde{V}^{54})/M_{\Delta}
\ee
denote the relevant weight coefficients in the $16_{F}^{i}\oplus 10_{F}$ space. The RH components of the three light states ($d^{ci}_{L}$) live in the orthogonal subspace defined by the relevant $4\times 4$ unitary transformation
\be\label{Ud}
\left(
\begin{array}{c}
d_{L}^{c1} \\
d_{L}^{c2} \\
d_{L}^{c3} \\
\hline
\tilde\Delta_{L}^{c}
\end{array}
\right)
=
\left(
\begin{array}{ccc|c}
A_d^{11} & A_d^{12} & A_d^{13} & B_d^{1} \\
A_d^{21} & A_d^{22} & A_d^{23} & B_d^{2} \\
A_d^{31} & A_d^{32} & A_d^{33} & B_d^{3} \\
\hline
C_d^{1} & C_d^{2} & C_d^{3} & D_d^{}
\end{array}
\right)
\left(
\begin{array}{c}
D_{L}^{c1} \\
D_{L}^{c2} \\
D_{L}^{c3} \\
\hline
\Delta_{L}^{c}
\end{array}
\right)
\ee 
The $A^{ij}_{d}$ and $B_{d}^{i}$ coefficients are constrained only from unitarity, and there is a lot of ambiguity in this sector. 
Introducing a compact notation
\be\label{Udcomp}
U_{d}\equiv
{\left(
\begin{array}{c|c}
A_{d} & \vec{B_{d}} \\
\hline
\vec{C_{d}}^{T} & D_{d}
\end{array}
\right)}
\ee 
the defining basis down-quark fields can be recast in terms of the physical ones as follows: 
\be\label{downdefphys}
\left(
\begin{array}{c}
\vec{D}_{L}^{c} \\
\hline
\Delta_{L}^{c}
\end{array}
\right)
=
\left(
\begin{array}{c|c}
A_{d}^{\dagger} & \vec{C_{d}}^{*} \\
\hline
\vec{B_{d}}^{\dagger} & D_{d}^{*}
\end{array}
\right)\left(
\begin{array}{c}
\vec{d}_{L}^{c} \\
\hline
\tilde\Delta_{L}^{c}
\end{array}
\right)
\ee
The relevant piece of the Yukawa lagrangian then reads
\bea\label{lagr2}
{\cal L}_{d}\!\!\!&\ni&\!\!\!Y^{ij}D_{L}^{i}\left[(A^{\dagger}_{d})^{jk}d_{L}^{ck}+C_{d}^{j*}\tilde{\Delta}_{L}^{c}\right]v_{d}^{10} \\
&-&\!\!\!F^{i}D_{L}^{i}\left[B_{d}^{*k}d_{L}^{ck}+D_{d}^{*}\tilde{\Delta}_{L}^{c}\right]v_{d}^{16}+
M_{\Delta}\Delta_{L}\tilde{\Delta}^{c}_{L},\nn
\eea
and thus the down quark mass matrix (in the $\{D_{L}^{i},\Delta_{L}\}$, $\{d_{L}^{ck},\tilde{\Delta}^{c}_{L}\}$ bases) becomes block-diagonal with zero at the $\Delta_{L}d^{c}_{L}$ position. Since the subsequent left-handed (LH) rotation is suppressed by ${\cal O}(v/M_{GUT})$, the left-handed physical components $d_{L}^{i}$, $\tilde{\Delta}_{L}$ can be (at leading order) identified with the defining ones $d_{L}^{i}\equiv D_{L}^{i}$, $\tilde{\Delta}_{L}\equiv \Delta_{L}$.

With all this at hand, the effective mass matrix for the down quarks $d^{i}$ obeys:
\be\label{Mdi}
M_{d}^{ik}=Y^{ij}(A_{d}^{\dagger})^{jk}v_{d}^{10}-F^{i}B_{d}^{*k}v_{d}^{16}+{\cal O}\left(\frac{v^{2}}{M_{GUT}}\right),
\ee
while the heavy state $\tilde{\Delta}$ has a mass $M_{\Delta}$. Recall that the $A_{d}^{\dagger}$ matrix and the $\vec{B}^{*}_{d}$ vector are just (hermitean conjugates of) the upper left and upper right blocks of the unitarity transformation (\ref{Udcomp}), which can be partially determined from the lower left $\vec{C}_{d}$ and lower right $D_{d}$ components of $U_{d}$ from the unitarity conditions $A_{d}^{\dagger}\vec{B}_{d}=- \vec{C}_{d}^{*}D_{d}$ and $|\vec{C}_{d}|^{2}+|D_{d}|^{2}=1$.

\paragraph*{Charged leptons:} 

The situation in the charged lepton sector is analogous to the down-quarks with the relevant parameters equipped by a subscript $l$ instead of $d$.  Taking into account the similarity of $M_{d}$ and $M_{l}^{T}$ one obtains:  
\be\label{Mli}
M_{l}^{ik}=(A_{l}^{*})^{ij}Y^{jk}v_{d}^{10}-B_{l}^{*i}F^{k}v_{d}^{16}+{\cal O}\left(\frac{v^{2}}{M_{GUT}}\right),
\ee
where, as before, $A_{l}$ and $\vec{B}_{l}$ complement the relevant $\vec{C}_{l}$ and $D_{l}$ defined as
\bea\label{chlparams}
\vec{C}_{l}=\vec{F}V^{16}/M_{\Lambda},\quad
{D}_{l}=(M_{10}+\tfrac{3}{2}\lambda\tilde{V}^{54})/M_{\Lambda}
\eea
with
$M_{\Lambda}=\sqrt{F^{\dagger}F (V^{16})^{2}+(M_{10}+\tfrac{3}{2}\lambda\tilde{V}^{54})^{2}}$ denoting the mass of the GUT-scale state $\Lambda$.

\paragraph*{The seesaw for neutrinos:}
After some tedium deferred to Appendix \ref{Appseesaw}, the standard seesaw formalism yields
\be\label{seesaw}
M_{\nu}\doteq M_{\nu}^\mathrm{II}-D_{\nu} M_{M}^{-1}D_{\nu}^{T}
\ee 
with the effective type-II and Dirac mass matrices obeying
\bea
M_{\nu}^\mathrm{II}&\equiv& (\vec{B}^{*}_{l}\otimes \vec{B}^{*}_{l}) \mu_{+}+A_{l}^{*}\mu_{3} A_{l}^{\dagger} \label{typeII} \\
D_{\nu}&\equiv& A^{*}_{l}Y v_u^{10}+(\vec{\zeta}\, \otimes \vec{B}^{*}_{l}) v_{u}^{10}V^{\overline{16}}/\Lambda_{\zeta}\nn.
\eea
The matrix $A_{l}$ and the  vector $\vec{B}_{l}$ are the same parameters that enter the charged lepton sector analysis above, c.f. formula (\ref{Mli}) and the comments in the Appendix A.

As it was already mentioned, we shall assume that the first term in (\ref{typeII}), i.e. the renormalizable part of the type-II contribution associated to the $SU(2)_{L}$-triplet in $54_{H}$, dominates over the non-renormalizable $\mu_{3}$-piece as well as the type-I contributions in the formula (\ref{seesaw}). Notice, however, that in such a case $M_{\nu}^{\mathrm{II}}$ has only one nonzero eigenvalue and thus the non-renormalizable type-II and/or type-I corrections should account, at some level, for the second nonzero neutrino mass and thus can never be entirely neglected.

Let us finish this section with a brief recapitulation of the four sum-rules we have obtained so far:
\bea
\label{Mu}
M_{u}& = &Yv_{u}^{10}, \\
\label{Md}
M_{d}&= & YA_{d}^{\dagger}v_{d}^{10}-\vec{F}\otimes \vec{B}_{d}^{*}v_{d}^{16},\\
\label{Ml}
M_{l}& = & A_{l}^{*}Yv_{d}^{10}-\vec{B}_{l}^{*}\otimes \vec{F} v_{d}^{16},\\
\label{Mnu}
M_{\nu} & \propto &\vec{B}^{*}_{l}\otimes \vec{B}^{*}_{l}+\ldots
\eea
The proportionality sign $\propto$ in (\ref{Mnu}) reflects the fact that the overall scale of the (triplet driven) type-II dominated neutrino mass matrix is unknown and the $\otimes$ symbol in $(\vec{x}\otimes \vec{y})_{ij}\equiv x_{i} y_{j}$ represents the outer products of the vectors $\vec{F}$ and $\vec{B}_{d,l}$ in equations  (\ref{Mdi}) and (\ref{Mli}). The formulae (\ref{Mu})-(\ref{Mnu}) shall be the subject of a detailed analysis in the reminder of this work.   
\subsection{Physical understanding \& decoupling\label{physics}}
One can check the consistency of formulae (\ref{Mu})-(\ref{Mnu}) by exploring the various limiting cases where different intermediate symmetries should be restored and the corresponding effective mass sum-rules revealed.

\paragraph*{$V^{16}, V^{54}\ll M_{10}$:} This setting corresponds to decoupling of $10_{F}$, so $\vec{C}_{l,d}\to 0$, $D_{d,l}\to 1$ and (from unitarity of $U_{d,l}$, c.f. equation (\ref{Udcomp})) $\vec{B}_{d,l}\to 0$ so the $A_{d,l}$ matrices become unitary. 
The light spectra are sensitive only to the electroweak $SU(2)_{L}\otimes U(1)_{Y}$ breakdown, but there is no means to transfer therein the information about the $SU(5)$ or Pati-Salam breaking at the renormalizable level.
This,  as expected,  leads to degenerate spectra of $M_{d}=M_{l}^{T}\propto M_{u}$ \`a-la $SO(10)$ with a single ``Yukawa-active'' Higgs multiplet $10_{H}$.

\paragraph*{$V^{54}\ll V^{16}$ :} This scenario features an intermediate $SU(5)$ symmetry - though the CG coefficients in $D_{d}$ and $D_{l}$ remain ``visible'' for $M_{10}\lesssim V^{54}$, one still has  $M_{\Delta}\approx M_{\Lambda}$, $\vec{C}_{d}\approx\vec{C}_{l}\neq 0$ and $D_{d}\approx D_{l}\neq 1$. This gives $\vec{B}_{d}\approx \vec{B}_{l}$ and thus the subsequent $SU(5)$ breaking affects the down-quarks and charged leptons in the same manner, and we get $M_{d}=M_{l}^{T}$ along the $SU(5)$ lines. However, the spectra of the up- and down-type quarks are disentangled. Apart from the potential problem with the proton decay there is also no handle on the CKM mixing in this case.

\paragraph*{$V^{16}\ll V^{54}$, $M_{10}\lesssim V^{54}$:} In this regime $10_{F}$ does feel the $SU(5)$ breaking in $54_{H}$, but due to the weakness of its interaction with the matter spinors (suppressed by $V^{16}/M_{10}$ or $V^{16}/V^{54}$), it can not transmit the information to the light sector (because $\vec{C}_{l,d}\to 0$, $D_{d,l}\to 1$) and again $\vec{B}_{d,l}\to 0$. Moreover, $A_{d,l}$ become unitary, leading to the same shape of the effective matter spectrum as in the decoupling case $V^{16}, V^{54}\ll M_{10}$.

As already mentioned above, a potentially realistic scenario could arise only if $V^{16}\lesssim V^{54}$ and $M_{10}$ low enough not to screen the CG coefficients in $D_{d,l}$. In such a case, one gets a good reason for the smallness of the CKM mixing (because $V_{CKM}\to \mathbbm{1}$ for $V^{16}\ll V^{54}$) 
while the 1st and 2nd generation Yukawa degeneracy can be lifted. The proximity of the third generation Yukawas could be reconciled with the third generation mass hierarchies for large $\tan\beta$ values. Recall that, indeed, $V^{16}\lesssim V^{54}$ is also the hierarchy suggested by the SUSY gauge coupling unification.
 
\section{Analysis and discussion}
Let us now inspect in detail the sum-rules (\ref{Mu})-(\ref{Mnu}). Apart from the  $3\times 3$ Yukawa matrix $Y$ common to $M_{u}$, $M_{d}$ and $M_{l}$, the latter two contain extra factors $A_{d,l}$ and $\vec{B}_{d,l}$ arising upon integrating out the heavy sector.
How much do we actually know about the $A_{d,l}$ matrices and the $\vec{B}_{d,l}$ vectors given the couplings $Y$, $\vec{F}$, $\lambda$ and the high-scale parameters $M_{10}$, $V^{54}$ and $V^{16} $?

\subsection{General prerequisites\label{generals}}
All the information we have about these quantities comes from the unitary of the $U_{d,l}$ matrices (\ref{Ud}):
\be\label{unitarity}
A_{d,l}^{\dagger}\vec{B}_{d,l}=- \vec{C}_{d,l}^{*}D_{d,l},\qquad |\vec{C}_{d,l}|^{2}+|D_{d,l}|^{2}=1.
\ee 
Several comments are worth making at this point.
First, due to reality of $M_{\Delta,\Lambda}$, the phases in $\vec{C}_{d,l}$ are aligned, while those of $D_{d}$ and $D_{l}$ can differ.
Second, the physical observables (i.e. spectra and mixings) coming from  (\ref{Mu})-(\ref{Mnu}) should be blind to any (unphysical) change of basis in the light sector.
 
To check this explicitly, recall that a general $4\times 4$ unitary matrix $U$ (parametrized by 6 angles and 10 phases
) can be written as a product of a unitary $3\times 3$ matrix $U_{3}$ (which depends on 3 angles $\xi_{1,2,3}$ corresponding to rotations in the 2-3, 3-1 and 1-2 planes respectively and 6 phases $\rho_{1,\ldots,6}$), acting on the first three indices only, and a ``unitary remnant'' $U_{4}$ accounting for the remaining $3$ mixings $\alpha_{1}$, $\alpha_{2}$, $\alpha_{3}$, corresponding to rotations in the 1-4, 2-4 and 3-4 planes, plus the remaining 4 phases $\psi_{1,..,4}$). We get:
\be\label{Ugeneric}
U=\left(
\begin{array}{c|c}
U_{3}(\xi_{1,2,3};\rho_{1,\ldots,6})  & {0} \\
\hline
{0} & 1
\end{array}
\right) U_{4}(\alpha_{1,2,3}; \psi_{1,..,4}),
\ee
where
\be\label{U4generic}
U_{4}(\alpha_{1,2,3}; \psi_{1,..,4})\!\!= \!\!e^{i\psi_{4}}R_{14}(\alpha_{1},\!\psi_{1})R_{24}(\alpha_{2},\!\psi_{2})R_{34}(\alpha_{3},\!\psi_{3}).
\ee
The $R_{i4}(\alpha_{i},\psi_{i})$ matrices in equation (\ref{U4generic}), given by
\be
R_{i4}(\alpha_{i},\psi_{i})\!=\!\left(\!
\begin{array}{ccc|c}
1&. & . &. \\
. &\cos\alpha_{i}&.   & -\sin\alpha_{i}e^{i\psi_{i}} \\
. &. & 1 &. \\
\hline
. & \sin\alpha_{i}e^{-i\psi_{i}}  & . & \cos\alpha_{i} 
\end{array}
\!\!\right),
\ee
represent the elementary unitary transformations in the $i$-$4$ planes.
The point here is that if we employ the parametrization (\ref{Ugeneric}) for $U_{d,l}$, the lower two sub-blocks of such unitary matrices are simple functions of $\alpha_{1,2,3}$ and $\psi_{1,..,4}$ (thus leading to a convenient parametrization of $\vec{C}_{d,l}$ and $D_{d,l}$). Indeed, performing the multiplications in (\ref{Ugeneric}) and (\ref{U4generic}) one obtains:
\bea\label{CDdefinition}
\vec{C}^{T}&=&e^{i\psi_{4}}(\sm_{1}e^{-i\psi_{1}}, \cm_{1}\sm_{2}e^{-i\psi_{2}}, \cm_{1}\cm_{2}\sm_{3}e^{-i\psi_{3}}), \nn\\
D&=& e^{i\psi_{4}} \cm_{1}\cm_{2}\cm_{3},
\eea
where the standard shorthand notation $s_{i}\equiv \sin\alpha_{i}$, $c_{i}\equiv \cos\alpha_{i}$ has been used and all the flavour indices $d$, $l$ distinguishing among the down quark and charged lepton sector quantities were dropped for simplicity. 

The main benefit from (\ref{CDdefinition}) is the independence of $\vec{C}_{d,l}$ and $D_{d,l}$  on the parameters driving the ``unphysical'' $U_{3}^{d,l}$ rotations. In fact, $U_{3}$ enters only the formulae for the $A$-matrices obeying
$
A={U}_{3}V
$
(where the family indices are suppressed), where $V$ denotes the $3\times 3$ upper left block of  the $U_{4}$ matrix:
\be\label{Vdefinition}
V\equiv e^{i\psi_{4}}\left(
\begin{array}{ccc}
\cm_{1}& -e^{i(\psi_{1}-\psi_{2})}\sm_{1}\sm_{2}& -e^{i(\psi_{1}-\psi_{3})}\cm_{2}\sm_{1}\sm_{3} \\
0 &\cm_{2}&-e^{i(\psi_{2}-\psi_{3})}\sm_{2}\sm_{3} \\
0 &0 & \cm_{3}
\end{array}
\right).
\ee
Note that $V$ actually measures the non-unitarity of $A$, since  $V$ becomes unitary if and only if $\vec{C}\to 0$, $|D|\to 1$, i.e. in the decoupling limit $M_{10}\to \infty$. Note that it also depends only on the reduced set of parameters $\alpha_{1,2,3}$ and $\psi_{1,\ldots,4}$.

\subsection{Hiding ``unphysical'' parameters $\xi_{1,..,3}$ and $\rho_{1,..,6}$}
The parametrization introduced in section \ref{generals} allows for recasting the 
sum rules (\ref{Md}), (\ref{Ml}) for $M_{d}$ and $M_{l}$ 
as:
\bea
M_{d}A_{d} & = &  YA_{d}^{\dagger}A_{d}v_{d}^{10}-\vec{F}\otimes (A_{d}^{\dagger}\vec{B}_{d})^{*}v_{d}^{16}, \nn\\
M_l^{T}A_{l}& = & YA_{l}^{\dagger}A_{l}v_{d}^{10}-\vec{F}\otimes (A_{l}^{\dagger}\vec{B}_{l})^{*}v_{d}^{16},\label{simp1}
\eea
where the $U_{3}^{d,l}$ matrices cancel in $A_{d,l}^{\dagger}A_{d,l}=V_{d,l}^{\dagger}V_{d,l}$ (with $V_{d,l}$ of the generic form (\ref{Vdefinition})) and due to unitarity (\ref{unitarity}), the RH-side (RHS) of relation (\ref{simp1}) becomes $\xi_{1,2,3}$- and $\rho_{1,\ldots,6}$-independent. The only trace of the ${U}_{3}^{d,l}$ rotations remains in the $A_{d,l}$ on the left-hand side (LHS) of (\ref{simp1}), but this can be dealt with by multiplying (\ref{simp1}) with $V_{d,l}^{-1}$:
\bea\label{Ydphysparams}
 M_{d}{U}_{3}^{d}  & = &  YV_{d}^{\dagger}v_{d}^{10}+(\vec{F}\otimes \vec{C}_{d}) D_{d}^{*}V_{d}^{-1} v_{d}^{16}, \nn\\
\label{YlTphysparams}  {M_{l}}^{T}{U}_{3}^{l}  & = &  YV_{l}^{\dagger}v_{d}^{10}+(\vec{F}\otimes \vec{C}_{l}) D_{l}^{*}V_{l}^{-1} v_{d}^{16}.
\eea
The  unitary transformations ${U}_{3}^{d,l}$ pending on the LHS of (\ref{YlTphysparams}) can then be eliminated upon looking at quantities like LHS.LHS$^{\dagger}$ or by a suitable redefinition of $d_{R}$ or $l_{L}$, which of course does not affect $V_{CKM}$, but can be relevant for the lepton mixing. 

Concerning the type-II dominated neutrino sector, the relevant mass matrix in the basis we used for the charged lepton sum-rule (\ref{YlTphysparams}) 
reads
\be
M_{\nu}\propto V_{l}^{-1T}A_{l}^{T}\vec{B}^{*}_{l}\otimes \left(V_{l}^{-1T}A_{l}^{T}\vec{B}^{*}_{l}\right)^{T}
\ee
which (using the unitarity conditions (\ref{unitarity})) can be rewritten in the form (dropping the $D^{*2}_{l}$ factor due to the overall scale ambiguity):
\be\label{MnuII}
M_{\nu}\propto V_{l}^{-1T}(\vec{C}_{l}\otimes \vec{C}_{l}) V_{l}^{-1}.
\ee
As mentioned before, such a mass matrix has only 1 nonzero eigenvalue and must be clearly subject to subleading corrections coming from type-I sector in order to lift at least one of the two remaining neutrino masses. This means that the only piece of information one can derive from (\ref{MnuII}) is the mixing angle between the heaviest third and the lighter second generation $\theta_{23}^{l}$. Therefore, we shall not consider neutrinos in the $3\times 3$ analysis in section \ref{numerics}.  

\subsection{Physical parameter counting}
Apart from the 18  parameters ($\xi_{1,..,3}^{d,l}$ and $\rho_{1,..,6}^{d,l}$) hidden in the $U_{3}^{d,l}$ matrices in (\ref{Ydphysparams}), we can eliminate other 6 quantities  by exploiting the close connection between the $M_{d}$ and $M_{l}^{T}$ matrices (which are identical up to one CG coefficient, c.f. formulae (\ref{Mdproto}) and (\ref{Mlproto})). 

Since the $\psi_{4}^{d}$ and $\psi_{4}^{l}$ phases entering $\vec{C}_{d,l}$, $D_{d,l}$ and $V_{d,l}$ given by the generic formulae (\ref{CDdefinition}) and (\ref{Vdefinition}) act only as global rephasing on the RHS of (\ref{Ydphysparams}), they can be absorbed into the definition of ${U}_{3}^{d,l}$.
Denoting  $\vec{C}'_{d,l}\equiv e^{-i\psi_{4}^{d,l}}\vec{C}_{d,l}$, ${D}'_{d,l}\equiv e^{-i\psi_{4}^{d,l}}{D}_{d,l}$, ${V}'_{d,l}\equiv e^{-i\psi_{4}^{d,l}}{V}_{d,l}$ and  ${U}_{3}'^{d,l}\equiv e^{-i\psi_{4}^{d,l}}{U}_{3}^{d,l}$, one can rewrite equations (\ref{Ydphysparams}) as
\bea\label{Ydphysparams2}
M_{d}{U}_{3}'^{d} & = & YV_{d}'^{\dagger}v_{d}^{10}+(\vec{F}\otimes \vec{C}_{d}') D_{d}'V_{d}'^{-1} v_{d}^{16}, \nn\\
\label{YlTphysparams2} M_{l}^{T}{U}_{3}'^{l} & = & YV_{l}'^{\dagger}v_{d}^{10}+(\vec{F}\otimes \vec{C}_{l}') D_{l}'V_{l}'^{-1} v_{d}^{16},
\eea
where the reality of $D'_{d,l}$ has been used to drop the star.

Next, using (\ref{downparams}) and (\ref{chlparams}), one can connect the down-quark and the charged lepton sector $\vec{C}_{d,l}'$ vectors by means of a single real and positive  parameter $s\equiv M_{\Delta}/M_{\Lambda}$:
\be\label{sdef}
\vec{C}'_{l}= e^{-i\phi}s\, \vec{C}'_{d},
\ee
where $\phi\equiv \psi_{4}^{l}-\psi_{4}^{d}$ is the only physical remnant of the $\psi_{4}^{d,l}$ phases. Notice that formula (\ref{sdef}) together with (\ref{CDdefinition}) admits for trading all $\psi_{1,2,3}^{l}$ and $\alpha_{1,2,3}^{l}$ for $\psi_{1,2,3}^{d}$ and $\alpha_{1,2,3}^{d}$ with $s$ and $\phi$ only.
Furthermore, one can exploit the proportionality  (\ref{downparams}) to express also $\vec{F}$ in terms of this reduced set of parameters: $\vec{F}=e^{i\psi_{4}^{d}}M_{\Delta}/V^{16}\vec{C}_{d}'\equiv x\vec{C}_{d}'$. Subsequently, using $Y=M_{u}/v_{u}$ and defining 
$v_{d}^{10}\cot\beta/v_{d}\equiv r e^{i\psi_{r}}$, $x v_{d}^{16}e^{-i\psi_{r}}\equiv q e^{i\psi_{q}}$ and ${U}_{3}'^{d,l}e^{-i\psi_{r}}\equiv \tilde{U}_{d,l}$ one obtains:
\bea\label{Ydphysparams3}
M_{d}\tilde{U}_{d}& = & M_{u}V_{d}'^{\dagger}r+qe^{i\psi_{q}}(\vec{C}_{d}' \otimes \vec{C}_{d}' )D_{d}'V_{d}'^{-1}, \nn\\
\label{YlTphysparams3} {M_{l}}^{T}\tilde {U}_{l}& = & M_{u}V_{l}'^{\dagger}r+qe^{i\psi_{q}}(\vec{C}_{d}' \otimes \vec{C}_{l}') D_{l}'V_{l}'^{-1}.
\eea
The last trick consists in observing that the complicated structure of the $V'_{d,l}$ matrices (\ref{Vdefinition}) can be further simplified by means of (suppressing the flavour indices)\footnote{The point is that $\vec{C}'\otimes \vec{C}'$ is a projector to its only nonzero eigenvector which leads to a reduction of complexity.} :
\bea
(\vec{C}' \otimes \vec{C}' )V'^{-1}&=&(\vec{C}' \otimes \vec{C}' )N(\vec{C}')\\
V'^{\dagger}&=&[1-P({\vec{C}'})]N(\vec{C}')\nn
\eea
 where $N$ is a real and diagonal matrix function defined for a generic complex vector $\vec{z}$ by $N(\vec{z})\equiv\mathrm{diag}^{-1}(n_{1}, n_{2}, n_{3})$ with 
\bea
n_{1}&=&\sqrt{1-|z_{1}|^{2}},\label{Nmatrix}\\
n_{2}&=&\sqrt{1-|z_{1}|^{2}} \sqrt{1-|z_{1}|^{2}-|z_{2}|^{2}},\nn\\
n_{3}&=&\sqrt{1-|z_{1}|^{2}-|z_{2}|^{2}}\sqrt{1-|\vec{z}|^{2}},\nn
\eea
while  $P(\vec{z})$ obeys
\be
P(\vec{z})\equiv \left(
\begin{array}{ccc}
|z_{1}|^{2}& 0 & 0\\
z_{1}z_{2}^{*} &|z_{1}|^{2}+|z_{2}|^{2}& 0 \\
z_{1}z_{3}^{*} & z_{2}z_{3}^{*}& |\vec{z}|^{2}
\end{array}
\right).
\ee
Notice that unlike $V'$, it is trivial to invert a diagonal matrix to get $N$ from (\ref{Nmatrix}) and neither $N(\vec{z})$ nor $P(\vec{z})$ depends on the global phase of $\vec{z}$.
With this at hand, one can rewrite (\ref{MnuII}) and (\ref{YlTphysparams3}) into the final form
\bea\label{Ydphysparams4}
M_{d}\tilde{U}_{d} & = & \!\!\! \left\{M_{u}[1-P({\vec{f}})]r+ q(\vec{f} \otimes \vec{f})\sqrt{1-|\vec{f}|^{2}}\right\}\!N(\vec{f}), \nn\\
\label{YlTphysparams4} M_{l}^{T}\tilde{U}_{l}& = & \!\!\!\left\{M_{u}[1-P(s{\vec{f}})]r \right.  \\ 
& & \quad\;\;\qquad +\left. qse^{-i\phi}(\vec{f} \otimes \vec{f})\sqrt{1-|s\vec{f}|^{2}}\right\}\!N(s\vec{f}),\nn
\\
\label{Mnutensor}
M_{\nu}&\propto &N(s\vec{f})(\vec{f}\otimes \vec{f})N(s\vec{f}),
\eea
where $\vec{f}\equiv e^{i\psi_{q}/2}\vec{C}_{d}'$ has been used to absorb the remaining unphysical parameter $\psi_{q}$. 

\paragraph*{Parameter-counting:} apart from the up-quark masses in $M_{u}$, there is in total 3 angles $\alpha_{1,2,3}\equiv \alpha_{1,2,3}^{d}$ and three phases $\gamma_{1,2,3}\equiv 2\psi_{1,2,3}^{d}-\psi_{q}$ in the generic complex vector $\vec{f}$ entering (\ref{Ydphysparams4}) and (\ref{Mnutensor}):
\be
\vec{f}^{T}\equiv  (\sm_{1}e^{-i\frac{\gamma_{1}}{2}}, \cm_{1}\sm_{2}e^{-i\frac{\gamma_{2}}{2}}, \cm_{1}\cm_{2}\sm_{3}e^{-i\frac{\gamma_{3}}{2}}).
\ee
On top of that,  there are other four real parameters $q,r,s$ and $\phi$ on the RHS of (\ref{YlTphysparams4}) and (\ref{Mnutensor}), so altogether we are left with 10 free parameters to match the six eigenvalues of $M_{d}$ and $M_{l}$ and 4 CKM mixing parameters in the quark sector. 
Remarkably enough, there is a simple vocabulary that can be used to get the charged-lepton sum rule (\ref{Ydphysparams4}) out of the down-quark one:
\be\label{translation}
\vec{f}\to e^{-i\phi/2}s\,\vec{f},\quad q\to q/s.
\ee
We shall exploit this feature in the physical analysis in the next section.
\subsection{Extracting the physical information}
Given $M_{u}$, one can first exploit the sum-rule (\ref{Ydphysparams4}) to fit the three down quark masses and all the CKM parameters. For any set of values of $\alpha_{1,2,3}$, $\gamma_{1,2,3}$, $q$ and $r$, the RHS of eq. (\ref{Ydphysparams4}) (to be denoted by $R$) is fully specified. In the basis in which $M_{u}$ is diagonal, one can decompose $M_{d}\tilde{U}_{d}$ on the LHS of (\ref{Ydphysparams4}) as 
\be\label{CKMintroduced}
M_{d}\tilde{U}_{d}=V_{CKM}^{0}{\cal D}_{d}W=R,
\ee
where ${\cal D}_{d}$ is a diagonal form of $M_{d}$  and $V_{CKM}^{0}$ is a ``raw'' form
of the CKM matrix $V_{CKM}^{0}=P_{L}V_{CKM}P_{R}$ ($P_{L}\equiv \mathrm{diag}(e^{i\phi_{1}},e^{i\phi_{2}},e^{i\phi_{3}})$ and $P_{R}\equiv \mathrm{diag}(e^{i\phi_{4}},e^{i\phi_{5}},1)$ denote the phase factors necessary to bring $V_{CKM}$ into the standard PDG form \cite{Yao:2006px}), while $W$ represents a generic unitary right-handed rotation.  
One can first get rid of $W$ and $P_{R}$ by focusing on the combination $R.R^{\dagger}$:
\be\label{sqquarksumrule}
 P_{L}V_{CKM}|{\cal D}_{d}|^{2}V_{CKM}^{\dagger}P_{L}^{\dagger}=R.R^{\dagger}.
\ee

Second, the diagonal entries, the principal minors and the full determinant are insensitive to $P_{L}$ and, remarkably enough, some of these combinations can be further simplified.
Denoting 
$
d_{ij}\equiv (V_{CKM}|{\cal D}_{d}|^{2}V_{CKM}^{\dagger})_{ij}$, the equality of the diagonal elements in (\ref{sqquarksumrule}) yields 
\be
d_{ii}=r^{2}m_{u}^{i2}(1-g_{i})+ g_{i}q^{2}\Sigma g_{j}+ 2rq\cos\gamma_{i}m_{u}^{i}g_{i}\sqrt{1-\Sigma g_{j}}  \label{constraints1}
\ee
(no summation over $i$), where $g_{i}\equiv |f_{i}|^{2}$ (for $i=1,..,3$) are real numbers $\in \langle 0,1)$ and $m_{u}^{i}$ correspond to the relevant up-type quark masses.
On the other hand, the three main minors $\Delta_{i<j}\equiv d_{ii}d_{jj}-d_{ij}d_{ji}= d_{ii}d_{jj}-|d_{ij}|^{2}
$ obey
 \bea
\Delta_{i<j} &  = & -m_{u}^{i2}m_{u}^{j2}r^{4}+r^{2}\left(d_{ii}m_{u}^{j2}+d_{jj}m_{u}^{i2}\right) \label{constraints2} \\
&-&r^{2}q^{2}g_ig_j\!\! \left[m_{u}^{i2}+m_{u}^{j2}-2m_{u}^{i}m_{u}^{j}\cos (\gamma_{i}-\gamma_{j})\right]
 \nn.
 \eea
It is crucial that $d_{ii}$ and $\Delta_{i<j}$ depend only on the  physical quark sector data so the 6 relatively simple constraints (\ref{constraints1}) and (\ref{constraints2}) can be (at least in principle) used to solve for 6 out of the 8 unknown quark sector parameters $g_{i}$, $\gamma_{i}$, $r$ and $q$, given $m_{u}^{i}$, $m_{d}^{i}$ and $V_{CKM}$.

For example, from (\ref{constraints2}) one readily gets three independent combinations 
$g_{i}g_{j}, i\neq j$  
\be\label{gigj}
g_{i}g_{j}={b_{ij}}/{q^{2}},
\ee
where the $b_{ij}$ coefficients defined as
\be
b_{ij}\equiv
\frac{|d_{ij}|^{2}-\left(d_{ii}-r^{2}m_{u}^{i2}\right)\left(d_{jj}-r^{2}m_{u}^{j2}\right)}{r^{2}\left[m_{u}^{i2}+m_{u}^{j2}-2m_{u}^{i}m_{u}^{j}\cos (\gamma_{i}-\gamma_{j})\right]}\equiv\frac{b^{(n)}_{ij}}{b^{(d)}_{ij}}
\ee
depend on $r^{2}$ and $\gamma_{i}$ only. From (\ref{gigj}), the individual $g_{i}$'s are then given by
\be\label{gissolved}
g_{i}=\frac{1}{|q|}\sqrt{\frac{b_{ij}b_{ik}}{b_{jk}}},\quad (i\neq j\neq k \neq i) .
\ee
These quantities may be used in (\ref{constraints1}) to recast the $\gamma_{i}$ phases as functions of $r$ and $q$ which become the only pending quark sector parameters.
Recall also that consistency of equations (\ref{gigj}) and (\ref{gissolved})  requires $r$ and $q$ such that 
$g_{i}\in \left\langle 0,1\right)$ and $g_{i}g_{j}\in \left\langle 0,\tfrac{1}{4}\right)$ for $i\neq j$.

Concerning the charged lepton sector, there is no analogue of decomposition  (\ref{CKMintroduced}) because the neutrino mass matrix remains unconstrained. All we can write is
$M_{l}^{T}\tilde{U}_{l}=V^{L}_{l}{\cal D}_{l}V^{R\dagger}_{l}
$,
where $V_{l}^{L,R}$ are unitary diagonalization matrices without immediate physical significance.
Nevertheless, one can look at the spectrum of $M_{l}$ by means of the three basic invariants - the trace, the sum of the main minors and  the determinant of $M_{l}.M_{l}^{\dagger}$. 
Moreover, the right-hand sides of the trace and sum-of-the-minors formulae can be obtained from (the sum of) the right-hand-sides of \eqs{constraints1}{constraints2} upon replacing 
\be\label{table2}
g_{i}\to s^{2}g_{i}, \;\; q\to q/s\;\; \mathrm{and}\;\; \gamma_{i}\to \gamma_{i}+\phi,
\ee
which is just the vocabulary (\ref{translation}) rewritten for $g_{i}$ and $\gamma_{i}$. 

\subsection{Numerical analysis\label{numerics}}
Prior getting to the full-featured three generation fit, let us inspect in brief the basic features of the $2\times2$ case focusing on the second and third generation of quarks and leptons. The reason is that in such a case a further constraint on one lepton mixing angle can be derived from the type-II dominated seesaw formula (\ref{typeII}).
One can then expect the higher order corrections coming from the effective operators (or further vector multiplets above the GUT-scale) to account for the structure of the light sector. However, the effects of $10_{F}$ must be compatible with the 2nd and 3rd generation spectra already at this level, should the current approach be viable at all.
\subsubsection*{$2\times2$ heavy charged sector analysis} 
Forgetting for a while about the first row and column in the matrix relations (\ref{YlTphysparams4}) (which is technically achieved by $\alpha_{1}\to 0$ yielding also $g_{1}=0$, with $\gamma_{1}$ left unconstrained), the formulae (\ref{constraints1}) and (\ref{constraints2}) are affected accordingly. Denoting 
$
d^{(2)}_{ij}\equiv 
(V^{(2)}_{CKM}
|{\cal D}_{d}^{(2)}|^{2}V^{(2)\dagger}_{CKM})_{ij}
$
where 
$$
{\cal D}_{d}^{(2)}\equiv \mathrm{diag}(m_{s},m_{b}),\quad 
V^{(2)}_{CKM}\equiv\left(
   \begin{array}{cc} 
      \cos \theta_{23}^{q} & \sin \theta_{23}^{q} \\
        -\sin \theta_{23}^{q} & \cos \theta_{23}^{q}
   \end{array}
\right)
$$
are the 23-blocks of ${\cal D}_{d}$ and $V_{CKM}$, one arrives at:  
\bea
d^{(2)}_{22} & = & r^{2}m_{c}^{2}(1-g_{2})  + g_2q^{2}(g_2+g_3)\nn \\
&+ &2rq\cos\gamma_{2}m_{c}g_{2}\sqrt{1-g_2-g_3}\label{EQQS1},\\
d^{(2)}_{33} &= & r^{2}m_{t}^{2}(1-g_3) + g_3q^{2}(g_2+g_3) \nn \\
&+& 2rq\cos\gamma_{3}m_{t}g_3\sqrt{1-g_2-g_3}\label{EQQS2}, 
\\
\mathrm{det}[d^{(2)}]  & = & -m_{c}^{2}m_{t}^{2}r^{4}+r^{2}\left(d_{22}^{(2)}m_{t}^{2}+d^{(2)}_{33}m_{c}^{2}\right)  \label{EQQS3}\nn
\\
&-&r^{2}q^{2}g_2g_3 \left[m_{c}^{2}+m_{t}^{2}-2m_{c}m_{t}\cos (\gamma_{2}-\gamma_{3})\right] \nn.
\eea
The corresponding lepton sector relations can be derived from $\Delta_{2<3}$ in (\ref{constraints2}) and the sum\footnote{Recall the individual diagonal entries of $M_{l}$ are unknown due to the RH-rotation ambiguity in $M_{l}^{T}\tilde{U}_{l}=V^{L}_{l}{\cal D}_{l}V^{R\dagger}_{l}$ while the trace remains fixed.} of equations (\ref{EQQS1}) and 
(\ref{EQQS2}) using the vocabulary (\ref{table2}):
\bea \label{EQLS1} 
m_{\mu}^{2}+m_{\tau}^{2}  & = & 
\left(g_2+g_3\right)^{2}s^{2}q^{2} 
+r^{2}\left[m_{c}^{2}\left(1-s^{2}g_2\right)\right.
\\
&  
+ &\left.m_{t}^{2}\left(1-s^{2}g_3\right)\right] + 2rs q\sqrt{1-s^{2}(g_2+g_3)}\times \nn\\
&  \times &  \left[g_2m_{c}\cos{(\gamma_{2}+\phi)}+g_3m_{t}\cos{(\gamma_{3}+\phi)}\right],\nn
\eea
\bea
m_{\mu}^{2}m_{\tau}^{2} & = & \Bigl\{r^{2} m_{t}^{2}m_{c}^{2}\left[1-\left(g_2+g_3\right)s^{2}\right] +s^{2}q^{2}\left[m_{t}^{2}(g_2)^{2}\right. \nn\\
& + & \left.m_{c}^{2}(g_3)^{2}\right] + 2m_{c}m_{t}s q\Bigl[g_2g_3s q\cos2(\gamma_{2}-\gamma_{3})  \label{EQLS2} \nn \\
& + & r\sqrt{1-s^{2}(g_2+g_3)}\times  \\
& \times & \left(g_2m_{t}\cos{(\gamma_{2}-\phi)}+g_3m_{c}\cos{(\gamma_{3}-\phi)}\right)\Bigr]\Bigr\} r^{2}. \nn
\eea
Formulae (\ref{EQQS1})-(\ref{EQLS2}) allow for a full reconstruction of the 5 relevant measurables,  (apart from $m_{t}$ and $m_{c}$ that we count amongst inputs) namely  $m_{s}$, $m_{b}$, $m_{\mu}$, $m_{\tau}$ and the 23 CKM mixing angle $\theta^{q}_{23}$ (recall that there is no CP phase in the $2\times 2$ quark sector) in terms of 5 real parameters ($q$, $r$, $s$, $\alpha_{2}$ and $\alpha_{3}$) and 3 phases ($\gamma_{2}$, $\gamma_{3}$, $\phi$). 

Thus, it is natural to constrain the fit furthermore by sticking to the CP-conserving (i.e. real) case, which corresponds to 0 or $\pi$ of the $\gamma_{1,2}$ and $\phi$ phases. This, however, makes the fit nontrivial because $\alpha_{1,2}\in \langle 0,2\pi)$ live in a compact domain, $s$ must be an ${\cal O}(1)$ number (to avoid extra fine-tuning in $M_{\Delta,\Lambda}$) and $q$ should (for consistency reasons\footnote{While the third family hierarchy should be compensated by a suitable choice of $r$, $q$ governs the second family scales and thus (in order to have $g_{i}q$ in (\ref{EQLS2}) around $m_{c}$ for $g_{i}\approx 10^{-1}$) $q$ should be within a few GeV range.}) be within a few GeV range. Thus, the only free parameter in the real $2\times2$ case is $r$.

Remarkably enough, even such a constrained setting admits good fits of all the relevant experimental data. A pair of illustrative solutions is given in TABLE \ref{TabSample2x2solution}.
One can see that $r$ plays the role of the $m_{t}/m_{b}$ hierarchy ``compensator'', while all the other parameters fall into their proper domains specified above. The relative smallness of $g_{2,3}$ indicates that we are indeed in the $V^{16}\lesssim V^{54}$ regime, as suggested by the qualitative arguments 
in section \ref{physics}. 
\begin{table}[h]
\begin{center}
\begin{tabular}{|c|c|c|c|c|c|c|}
\cline{1-3} \cline{5-7}
\multicolumn{3}{|c|}{Sample solution 1} & \,\,& \multicolumn{3}{|c|}{Sample solution 2}  \\
\cline{1-3} \cline{5-7}
parameter & value & deviation & &  parameter & value & deviation\\
\cline{1-3} \cline{5-7}
\multicolumn{3}{|c|}{Input} & & \multicolumn{3}{|c|}{Input}  \\
\cline{1-3} \cline{5-7}
$\gamma_{2}$ & $\pi$ &-& & $\gamma_{2}$ & $\pi$ &- \\
$\gamma_{3}$ & $\pi$ & -& &$\gamma_{3}$ & $\pi$ &- \\
$\phi$ & $\pi$ & -& & $\phi$ & $\pi$ & - \\
$m_{c}$ [GeV] & $0.209$ & c.value & & $m_{c}$ [GeV]  &$0.209$ &  c.value\\
$m_{t}$ [GeV] & $90$ & c.value& & $m_{t}$ [GeV] & 70 & $\sim 1\sigma$  \\
\cline{1-3} \cline{5-7}
\multicolumn{3}{|c|}{Free parameters} & & \multicolumn{3}{|c|}{Free parameters}  \\
\cline{1-3} \cline{5-7}
$r$ & $0.0143$ & - & &  $r$ & 0.0184 & -   \\
$s$ & $2.3825$ & -& & $s$ & 2.2968& -   \\
$g_{2}$ & $0.0187$ & -& & $g_{2}$ & 0.0199 & -  \\
$g_{3}$ & $0.0415$ & -& & $g_{3}$ & 0.0414& -  \\
$q$ [GeV] & $1.6296$ &- & & $q$ [GeV] & 1.5787& -  \\
\cline{1-3} \cline{5-7}
\multicolumn{3}{|c|}{Output} & & \multicolumn{3}{|c|}{Output}  \\
\cline{1-3} \cline{5-7}
$m_{s}$ [GeV] & $0.0299$ & c.value & & $m_{s}$ [GeV]  & $0.0299$ & c.value \\
$m_{b}$ [GeV] & $1.200$ & $\sim 1\sigma$ & & $m_{b}$ [GeV]& $1.200$ & $\sim 1\sigma$    \\
$m_{\mu}$ [GeV] & $0.0756$ & c.value  & &$m_{\mu}$ [GeV]& $0.0756$ & c.value\\
$m_{\tau}$ [GeV] & $1.292$ & c.value  & &$m_{\tau}$ [GeV] & $1.292$ & c.value\\
$\sin\theta^{q}_{23}$ & $0.036$ & c.value & & $\sin\theta^{q}_{23}$ & $0.036$ & c.value\\
\cline{1-3} \cline{5-7}
\end{tabular}
\end{center}
\caption{\label{TabSample2x2solution} A pair of illustrative examples of the 2$\times$2 fits (i.e. focusing on the second and third generation) in the real setting (i.e. all phases set to  $0$ or $\pi$). 
A sample set of GUT-scale inputs was taken from \cite{Das:2000uk} for $\tan\beta\approx 10$.}
\end{table}
\subsubsection*{Large lepton mixing  in the CP-conserving $2\times2$ case} 
In the $2\times 2$ case, one can extend the current analysis to the neutrino sector because the $2\times 2$ version of formula (\ref{Mnutensor}) can be a good leading order neutrino mass matrix contribution (for hierarchical case). Sticking again to the CP conserving setting, one can rewrite the core of formula (\ref{Mnutensor}) (taking for simplicity $\gamma_{2}=\gamma_{3}$, that accounts only for an irrelevant overall sign) in the form:
\be
\vec{{f}}\otimes \vec{f}=
\left(
   \begin{array}{cc} 
      g_{2} & \sqrt{g_{2}g_{3}} \\
       . & g_{3}
   \end{array}
\right) 
\ee 
and
$$
N(s\vec{{f}})=\mathrm{diag}
\left(
      \sqrt{1-s^{2}g_{2}},
        \sqrt{1-s^{2}(g_{2}+g_{3})}
\right)^{-1},
$$
which in the regime suggested by the charged lepton fit ($s^{2}g_{i}\ll 1$) leads to an approximate formula\footnote{The charged lepton contribution to the lepton mixing is negligible because of the hierarchical nature of the LHS of eq. (\ref{YlTphysparams4}).}
\be\label{thetageneral}
\tan 2\theta_{23}^{l}\approx 2{\sqrt{g_{2}g_{3}}}/{|g_{2}-g_{3}|},
\ee\nopagebreak
so the proximity of $g_{2}$ and $g_{3}$ leads to a large 2-3 lepton sector mixing! Numerically, for $g_{2}\approx 0.0187$ and $g_{2}\approx 0.0415$ (solution 1 in TABLE \ref{TabSample2x2solution}) one gets $\sin^{2}\!2\theta_{23}^{l}\approx 0.85$, which is remarkably close to the observed nearly maximal atmospheric mixing \cite{Strumia:2006db}.

It is interesting that the $g_{2}\sim g_{3}$ case is not accidental. Notice first that (neglecting the small $P(\vec{f})\ll 1$ and the almost unity matrices $N(\vec{f})$ and $\tilde{U}_{d}$), the down-quark mass sum-rule in (\ref{Ydphysparams4}) reads at leading order:
\be\label{MDapprox}
M_{d}\approx \left(
   \begin{array}{cc} 
      m_{c} & 0 \\
        0 & m_{t}
   \end{array}
\right) r+q \left(
   \begin{array}{cc} 
      g_{2} & \sqrt{g_{2}g_{3}} \\
        \sqrt{g_{2}g_{3}} & g_{3}
   \end{array}
\right).
\ee
The hierarchies in TABLE \ref{TabSample2x2solution} suggest that the second term in (\ref{MDapprox}) dominates over the first one in all but the 22 entry. Thus, we have approximately:
\be
M_{d}\approx \left(
   \begin{array}{cc} 
      q g_{2} & q \sqrt{g_{2}g_{3}} \\
        q \sqrt{g_{2}g_{3}} & r m_{t}
   \end{array}
\right)\Rightarrow\quad m_{s}\approx q g_{2},\; m_{b}\approx r m_{t}.
\ee
Since $M_{u}$ is diagonal, the 23 quark sector mixing angle $\theta^{q}_{23}$ comes entirely from $M_{d}$, i.e. 
$
V_{cb}\approx \sin\theta^{q}_{23}
\approx q {\sqrt{g_{2}g_{3}}}/{m_{b}}.
$ 
Solving for $g_{2}$ and $g_{3}$ and substituting into (\ref{thetageneral}) one finally obtains
 \be
\tan 2\theta_{23}^{l}\approx 2 x / \left|1-x^{2}\right|\quad \mathrm{where}\quad  x\equiv \frac{y_{b}}{y_{s}}\sin\theta^{q}_{23},
\ee
in agreement with the numerical example given above.
 
Therefore, we have obtained an interesting correlation between $x\sim 1$ (i.e. the proximity of $V_{cb}$
to $y_{s}/y_{b}$) in the quark sector and the large atmospheric mixing $\theta_{23}^{l}$ for the type-II dominated neutrino mass matrix, quite along the lines of the connection of the $\theta_{23}^{l}$-maximality and the $b-\tau$ Yukawa convergence in the context of SUSY $SO(10)$ models with $\overline{126}_{H}$ in the Higgs sector \cite{b-taularge23}.
\subsubsection*{Full-featured $3\times3$ charged sector analysis} 
Though the full three-generation case is much more sensitive to all sorts of higher order corrections it, can  still be instructive to look at the fit of the charged sector formulae (\ref{constraints1}) and (\ref{constraints2}). Perhaps the simplest approach would be to perturb the $2\times 2$ fits studied in the previous section by relaxing the conditions imposed on $\alpha_{1}$, $\gamma_{2,3}$ and $\phi$ and admitting slight changes in the other parameters as well, in order to fit the first generation quantities $m_{d}$, $\theta_{12}^{q}$, $\theta_{13}^{q}$, $\delta$ and $m_{u}$. 

It is very interesting that this simple strategy fails due to the tension emerging already at the level of the pure quark sector fit, which (being even underconstrained) should be essentially trivial. In particular, we found that the quark sector data are reproduced only for the price of pushing the $r$-parameter very far from its natural domain $r\approx m_{b}/m_{t}\approx 10^{-2}$ suggested by the $2\times 2$ fit, c.f. TABLE \ref{TabSample2x2solution}. Instead, all the numerical quark sector fits we found give $r\approx m_{s}/m_{c}\approx 10^{-1}$ and a $g_{3}$ value very close to 1 indicating a high degree of fine-tuning in formula (\ref{EQQS1}). An interested reader can find a sample set of relevant data in TABLE \ref{TabSample3x3solution}.
\begin{table}[h]
\begin{center}
\begin{tabular}{|c|c|cc|c|}
\cline{1-2}
\multicolumn{2}{|c|}{Physical inputs} &\;\;\;\;\; & \multicolumn{2}{c}{} \\  
\cline{1-2}
$m_{u}$ [MeV]  & 0.720 & & \multicolumn{2}{c}{} \\ 
$m_{c}$ [GeV]  & 0.209 & & \multicolumn{2}{c}{} \\ 
$m_{t}$ [GeV]  & 90 & & \multicolumn{2}{c}{} \\
\cline{1-2}\cline{4-5}
\multicolumn{2}{|c|}{Free parameters} & & \multicolumn{2}{|c|}{Physical outputs} \\
\cline{1-2}\cline{4-5}
$\gamma_{1}$ & 2.122 & &  \multicolumn{1}{|c|}{ $m_{d}$ [MeV]}&1.562 \\
$\gamma_{2}$ & 1.725& & \multicolumn{1}{|c|}{ $m_{s}$ [GeV]}  & 0.030\\
$\gamma_{3}$ & 3.795 & & \multicolumn{1}{|c|}{$m_{b}$ [GeV]}  & 1.090\\
$r$ & 0.1527 & & \multicolumn{1}{|c|}{$\sin\theta^{q}_{12}$} & 0.2229 \\
$g_{1}$ & 0.0002 & & \multicolumn{1}{|c|}{$\sin\theta^{q}_{13}$} & 0.0365\\
$g_{2}$ & 0.0048 & & \multicolumn{1}{|c|}{$\sin\theta^{q}_{23}$} & 0.0032\\
$g_{3}$ & 0.9945 & & \multicolumn{1}{|c|}{$\delta_{CKM}$} & $60^{o}$ \\
\cline{4-5}
$q$ & 0.5436 & & \multicolumn{2}{c}{} \\
\cline{1-2}
\end{tabular}
\end{center}
\caption{\label{TabSample3x3solution} An example exhibiting the generic features of all the $3\times3$ quark sector fits we found. No $\chi^{2}$ is given because the number of free parameters exceeds the number of observables in the quark sector and the fit is in principle (though not technically) trivial.  Notice namely the value of $r$ being one order of magnitude higher than in the $2\times 2$ case (c.f. TABLE \ref{TabSample2x2solution}) as well as the singular behavior of $g_{3}$. As before, the sample set of inputs was taken from \cite{Das:2000uk} for $\tan\beta\approx 10$.}
\end{table}

This, however, has dramatic consequences for the charged lepton sector. With $r\sim 10^{-1}$ incapable of compensating the $m_{t}/m_{b}$ hierarchy without an extra aid from $q$, there is not enough freedom left to account for the $m_{\mu}/m_{s}$ hierarchy and  (for a fixed $m_{\mu}$) $m_{s}$ turns out to be generically too large.

One can use the formula (\ref{gigj}) to understand this peculiar instability analytically. 
Notice first that the denominators ${b^{(d)}_{ij}}$ of all the $b_{ij}$ coefficients in (\ref{gigj}) are always positive, so in order to have $g_{i}g_{j}\geq 0$ for all $i\neq j$, all the numerators ${b^{(n)}_{ij}}$ must be positive as well. 

\begin{figure}[t]
\epsfig{file=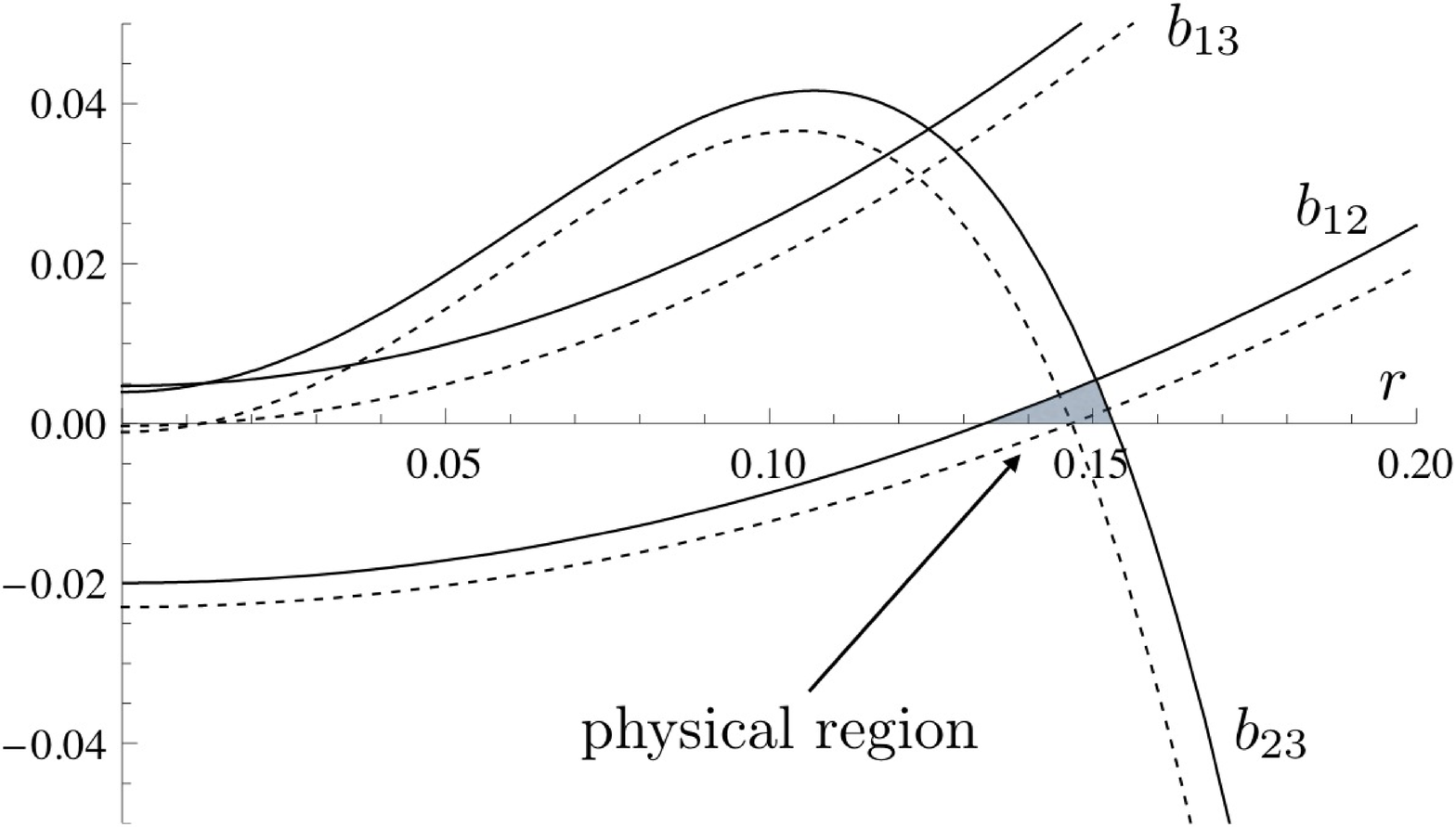 ,width=6.8cm}
\caption{\label{figure_curves} 
The qualitative $r$-behavior of numerators of the $b_{ij}$ coefficients in the $V_{CKM}=\mathbbm{1}$ limit (dashed curve) and in the physical $V_{CKM}\neq \mathbbm{1}$ setting (plain). Apart from the two common roots of the dashed curves at $r\approx 0.015$ and $r\approx 0.15$ there is a third one outside the displayed region at $r\approx 2.2$. The physical requirement $b_{ij}>0$ (for all $i \neq j$) can be satisfied only if $V_{CKM}\neq \mathbbm{1}$ and $r\approx 0.15$, which leads to an intrinsic instability in the quark sector fits, c.f. formula (\ref{numer2}) and discussion around. For optical reasons, $b_{12}$ was magnified by a multiplicative factor of $10^{7}$ while $b_{13}$ by $10^{2}$.}
\end{figure}

Suppose first that the CKM mixing can be neglected, i.e. $d_{ij}=(V_{CKM}|{\cal D}_{d}|^{2}V_{CKM}^{\dagger})_{ij}=m_{d}^{i2}\delta_{ij}$,  (no summation over $i$). The numerators $b^{(n)}_{ij}$ in such a case read:
\be\label{numer1}
b^{(n)}_{ij}(r)= - p_{i}(r)p_{j}(r),
\ee
where $p_{i}(r)=m_{d}^{i2}-r^{2}m_{u}^{i2}$ are simple polynomials in $r$. However,  given (\ref{numer1}), there is no\footnote{At least one product of two out of any three real numbers is always non-negative!} $r$ that would lead to $g_{i}g_{j}\geq 0$ for all $i\neq j$ (except $p_{i}(r)=0$ for at least one $i$ which, however, corresponds to $g_{i}=0$, c.f. (\ref{gissolved}), taking us back to the $2\times 2$ case).  Graphically, the three numerators $b^{(n)}_{ij}(r)$ correspond to three functions sharing roots on the real axis, as depicted by the dashed curves in FIG.~\ref{figure_curves}. 
This means that the $3\times 3$ quark spectra  (regardless of their particular shape) can not be accommodated in the simplest renormalizable model  {\it unless $V_{CKM}\neq \mathbbm{1}$}.

However, once the CKM mixing is turned on, the $p_{i}(r)$ polynomials change to $p'_{i}(r)=d_{ii}-r^{2}m_{u}^{i2}$ and an extra positive term shows up in the relevant analogue of formula (\ref{numer1}), c.f. also equation (\ref{gigj}):
\be\label{numer2}
{b^{(n)}_{ij}}'(r)=  |d_{ij}|^{2}-p'_{i}(r)p'_{j}(r),
\ee
with the net effect of slightly distorting and lifting the three dashed curves corresponding to the $V_{CKM}=\mathbbm{1}$ case. As a consequence, a small ``physical'' window for $r$ can open around the point where two of the original $b^{(n)}_{ij}(r)$ functions shared a root (i.e. around the roots of $p_{i}(r)$'s) while the third was positive, see again FIG.~\ref{figure_curves}.
Unfortunately, the value of $r$ where this happens corresponds to $r\approx m_{s}/m_{c}\approx 0.15$ and not $r\approx m_{b}/m_{t}\approx 0.015$ that would be compatible with the ``natural'' solution we have obtained in the $2\times2$ case.

Therefore, a single extra $10_{F}$ in the matter sector, and in particular its non-decoupling effects (that have been shown to account for all the quark and lepton masses and mixings in the two-generation case), can not provide the only source of physics contributing to the first generation observables. 

\section{Conclusions}
In this paper, we have scrutinized the effective Yukawa sector emerging in a class of renormalizable SUSY $SO(10)$ GUT models with $16_{H}\oplus\overline{16}_{H}$ Higgs fields driving the $SU(2)_{R}\otimes U(1)_{B-L}$ breakdown. 

An extra $SO(10)$-vector matter multiplet $10_{F}$ with an accidentally small singlet mass term (around the GUT scale) would not decouple from the GUT-scale physics and, under certain conditions, can provide a non-vanishing component of the light matter states (spanning in traditional case only on the three $SO(10)$ spinors $16_{F}^{i}$) through the mixing term $16_{F}^{i}10_{F}16_{H}$ in the superpotential. The sensitivity of $10_{F}$ to $SU(5)$ and $SU(2)_{R}\otimes U(1)_{B-L}$ breaking (through the $10_{F}10_{F}54_{H}$ and $16_{F}^{i}10_{F}16_{H}$ interactions) lifts the typical high degree of degeneracy in the effective low-energy Yukawa couplings, giving rise to a characteristic pattern of non-decoupling effects in the effective mass matrices. This, however, could render the Yukawa sector of the model potentially realistic. 
%

In order to deal with the complicated structure of the emerging effective matter sector mass sum-rules, a thorough analysis of the would-be ambiguities emerging upon integrating out the heavy parts of the matter spectra has been provided and the relevant parameter-counting was given. This admits for a detailed numerical analysis of the quark sector in the full three-generation case. 
If the renormalizable part of the type-II contribution associated to the $SU(2)_{L}$ triplet in $54_{H}$ governs the seesaw formula, the neutrino mass matrix becomes partly calculable. Focusing on the 2nd and 3rd generation, the 23-mixing (for hierarchical neutrino spectrum) can be estimated. In such a case, we found a striking  (GUT-scale) correlation between the proximity of $V_{cb}$ and $y_{s}/y_{b}$ and a large 23-mixing angle in the lepton sector: $\tan 2\theta_{23}^{l}\approx 2 x / \left|1-x^{2}\right|$ where  $x\equiv ({y_{b}}/{y_{s}})V_{cb}$.

Concerning the charged sector Yukawa sum-rules, any successful fit of the quark spectra in the simplest renormalizable scenario (with a single non-decoupling vector matter multiplet) requires a non-trivial CKM mixing $V_{CKM}\neq \mathbbm{1}$ and we provide a detailed analytical  understanding of this peculiarity. However, with the charged lepton sector spectrum taken into account, a generic tension in the $m_{\mu}/m_{s}$ hierarchy is revealed, calling for extra sources of corrections affecting the first generation observables, be it e.g. contributions from higer order operators or additional vector matter multiplets.

\section*{Acknowledgments}
I am grateful to Steve King for discussions throughout preparing this manuscript. The work was supported by the PPARC Rolling Grant PPA/G/S/2003/00096.
\appendix
\section{The seesaw\label{Appseesaw}}
Including all the effective contributions sketched in section \ref{GUTscalematrices}, the full $8\times 8$ neutrino mass matrix (in the $\{N_L,N_L^c,\Lambda_L^{0},\Lambda_L^{c0}\}$ basis) receives the following order of magnitude form (forgetting about the CG coefficients): 
\be\label{Mnuproto}
M_{\nu}
=\left(
\begin{array}{cccc}
\kappa({v_{u}^{\overline{16}})^{2}}/{\Lambda_{\kappa}} &  Y v_u^{10} &\vec{\zeta}v_{u}^{\overline{16}}v_{u}^{10}/{\Lambda}_{\zeta} & \vec{F} V^{16}  \\
 . & M_{M} & \vec{\zeta}\, V^{\overline{16}}v_{u}^{10}/\Lambda_{\zeta} & \vec{F} v_d^{16} \\
. &  . & \mu_{+} & M_{\Lambda\Lambda^{c}}\\
 . &  . & .& \mu_{-}
\end{array}\right)
\ee
with $M_{\Lambda\Lambda^{c}}\equiv M_{10}\!+\! \frac{3}{2}\lambda \tilde{V}^{54}$ and $\mu_{\pm}\equiv \lambda c_{T}w_\pm+\sigma(v_{u}^{10})^{2}/{\Lambda_{\sigma}}$.
The first row GUT-scale entries at the 14 position can be cancelled by a suitable rotation in the $N_{L}$-$\Lambda^{0}_{L}$ plane, identical to the charged sector $U_{l}$ transformation obtained from (\ref{Udcomp}) upon replacing $d\to l$ (recall the $SU(2)_{L}$-doublet nature of $L_{L}$ and $\Lambda_{L}$). 
Using the decompositions $N_{L}=A^{\dagger}_{l}n_{L}+\vec{B}^{*}_{l}\tilde{\Lambda}^{0}_{L}$ and $\Lambda_{L}^{0}=\vec{B}^{*}_{l}.n_{L}+D_{l}^{*}\tilde{\Lambda}^{0}_{L}$, which is just the lepton sector analogue of formula (\ref{downdefphys}), one obtains (in the $\{n_L, N_L^c,\tilde{\Lambda}_L^{0},\Lambda_L^{c0}\}$ basis):
\be
M_{\nu} =\left(
\begin{array}{c|ccc}
M_{\nu}^{\mathrm{II}} &  D_{\nu} &  \vec{B}^{*}_{l} D_{l}^{*}\mu_{+}& 0  \\
\hline
.& M_{M} & \vec{C}^{*T}Yv_{u}^{10} +  \vec{\zeta}\, V^{\overline{16}}D_{l}^{*}v_{u}^{10}/\Lambda_{\zeta} & \vec{F} v_d^{16} \\
. & . & D^{*}_{l}D^{*}_{l}\mu_{+} & M_\Lambda\\
  . & . & . & \mu_{-}
\end{array}\right)
\label{Mnuraw}
\ee
where $M_{\Lambda}$ is the lepton sector heavy state mass given by equation (\ref{chlparams}), $M_{\nu}^{\mathrm{II}}\equiv\vec{B}^{*}_{l}\otimes \vec{B}^{*}_{l}  \mu_{+}+A_{l}^{*}\mu_{3} A_{l}^{\dagger}$ and $D_{\nu}\equiv A^{*}_{l}Y v_u^{10}+(\vec{\zeta}\otimes \vec{B}^{*}_{l}) v_{u}^{10}V^{\overline{16}}/\Lambda_{\zeta}$.
The seesaw formula yields
\bea
M_{\nu}=M_{\nu}^{\mathrm{II}}- (D_{\nu},\vec{B}_{l}^{*}D_{l}^{*}\mu^{+},0)M_{234}^{-1}(D_{\nu},\vec{B}_{l}^{*}D_{l}^{*}\mu^{+},0)^{T}\nn\\
\label{largeseesaw}\eea
where $M_{234}$ is the  $\{N_{L}^{ci},\tilde{\Lambda}^{0}_{L},{\Lambda}^{c0}_{L} \}$ sector $5\times 5$ (i.e. lower-right) submatrix of (\ref{Mnuraw}). 
Denoting 
\be
m \equiv (\vec{C}^{*T}Yv_{u}^{10} + \vec{\zeta}D_{l}^{*}v_{u}^{10}\, V^{\overline{16}}/\Lambda_{\zeta}, F v_d^{16}) 
\ee
and 
$
 \tilde{M}_{\Lambda}\equiv
\left(\begin{array}{cc}
0 &  M_{\Lambda}  \\  
M_{\Lambda} & 0
\end{array}
\right)
$,
which stand  for the leading contributions in the upper-right ($m$) 3$\times$2 and lower-right ($m$) 2$\times$2 blocks of $M_{234}$ respectively, one can use $|m|\ll M_{\Lambda}, M_{M}$ to analytically invert $M_{234}$:
\be
M_{234}^{-1}\approx
\left(\begin{array}{cc}
M_{2}& -M_{M}^{-1}m\, {\tilde{M}_{\Lambda}}^{-1}\\  
-{\tilde{M}_{\Lambda}}^{-1}m^{T} M_{M}^{-1} & M_{34}
\end{array}
\right)
\ee
where $M_{2}\equiv M_{M}^{-1}+ M_{M}^{-1}\, m\, {\tilde{M}_{\Lambda}}^{-1} m^{T}M_{M}^{-1} $ and $M_{34}\equiv {\tilde{M}_{\Lambda}}^{-1}+ {\tilde{M}_{\Lambda}}^{-1}m^{T} {M_{M}}^{-1} m\, {\tilde{M}_{\Lambda}}^{-1}$, see \cite{Schechter:1981cv} for details.
The seesaw formula (\ref{largeseesaw}) then yields (\ref{seesaw})
up to higher order terms.

\end{document}